\documentclass[prb,twoside,twocolumn,superscriptaddress,showpacs,showkeys]{revtex4}
\usepackage{times}
\usepackage{amssymb,amsfonts,amsmath}
\usepackage{txfonts}
\usepackage{upgreek}
\usepackage{graphicx}
\usepackage{color}
\usepackage{units}

\begin{document}
\title{Anomalous insulator metal transition in boron nitride-graphene hybrid atomic layers}

\author{Li Song}
\affiliation{Department of Mechanical Engineering \& Materials Science, Rice University, Houston, TX 77005, USA}
\author{Luis Balicas}
\affiliation{National High Magnetic Field Lab, Tallahassee, FL 32310, USA}
\author{Duncan J. Mowbray}
\affiliation{Nano-Bio Spectroscopy Group and ETSF Scientific Development Centre, Departamento de F\'{\i}sica de Materiales, Universidad del Pa\'{\i}s Vasco, Centro de F\'{\i}sica de Materiales CSIC-UPV/EHU-MPC and DIPC, Avenida de Tolosa 72, E-20018 San Sebasti\'{a}n, Spain}
\author{Rodrigo B. Capaz}
\affiliation{Instituto de F\'{\i}sica, Universidade Federal do Rio de Janeiro, Caixa Postal 68528, Rio de Janeiro, RJ 21941-972, Brazil}
\affiliation{Department of Physics, University of California at Berkeley, Berkeley, California 94720, USA}
\author{Kevin Storr}
\affiliation{Department of Physics, Prairie view A\&M University, Prairie View, TX 77446, USA}
\author{Lijie Ci}
\affiliation{Department of Mechanical Engineering \& Materials Science, Rice University, Houston, TX 77005, USA}
\author{Deep Jariwala}
\affiliation{Department of Mechanical Engineering \& Materials Science, Rice University, Houston, TX 77005, USA}
\author{Stefan Kurth}
\affiliation{Nano-Bio Spectroscopy Group and ETSF Scientific Development Centre, Departamento de F\'{\i}sica de Materiales, Universidad del Pa\'{\i}s Vasco, Centro de F\'{\i}sica de Materiales CSIC-UPV/EHU-MPC and DIPC, Avenida de Tolosa 72, E-20018 San Sebasti\'{a}n, Spain}
\affiliation{IKERBASQUE, Basque Foundation for Science, E-48011 Bilbao, Spain}
\author{Steven G. Louie}
\email{sglouie@berkeley.edu}
\affiliation{Department of Physics, University of California at Berkeley, Berkeley, California 94720, USA}
\affiliation{Materials Sciences Division, Lawrence Berkeley National Laboratory, Berkeley, California 94720, USA}
\author{Angel Rubio}
\email{angel.rubio@ehu.es}
\affiliation{Nano-Bio Spectroscopy Group and ETSF Scientific Development Centre, Departamento de F\'{\i}sica de Materiales, Universidad del Pa\'{\i}s Vasco, Centro de F\'{\i}sica de Materiales CSIC-UPV/EHU-MPC and DIPC, Avenida de Tolosa 72, E-20018 San Sebasti\'{a}n, Spain}
\affiliation{Fritz-Haber-Institut der Max-Planck-Gesellschaft, Berlin, Germany}
\author{Pulickel M. Ajayan}
\email{ajayan@rice.edu}
\affiliation{Department of Mechanical Engineering \& Materials Science, Rice University, Houston, TX 77005, USA}



\begin{abstract}
The study of two-dimensional (2D) electronic systems is of great fundamental significance in physics. Atomic layers containing hybridized domains of graphene and hexagonal boron nitride (h-BNC) constitute a new kind of disordered 2D electronic system. Magneto-electric transport measurements performed at low temperature in vapor phase synthesized h-BNC atomic layers show a clear and anomalous transition from an insulating to a metallic behavior upon cooling. The observed insulator to metal transition can be modulated by electron and hole doping and by the application of an external magnetic field.  These results supported by ab-initio calculations suggest that this transition in h-BNC has distinctly different characteristics when compared to other 2D electron systems and is the result of the coexistence between two distinct mechanisms, namely, percolation through metallic graphene networks and hopping conduction between edge states on randomly distributed insulating h-BN domains.
\end{abstract}
\pacs{71.30.+h, 72.80.Vp, 73.22.Pr, 73.61.wp, 81.05.Uf}

\keywords{metal to insulator transition | graphene | boron nitride | h-BNC}

\maketitle

\section{Introduction}

The interplay between Coulomb interactions and randomness has been a long-standing problem in the physics of 2D electronic systems \cite{[1],[2]}. According to the scaling theory of localization, carriers in 2D systems become localized due to the presence of randomness (e.g. defects or impurities), causing the electrical resistance to increase as the temperature is lowered leading to an insulating ground state. Consequently, the non-existence of a metallic state or the non-occurrence of an insulator to metal transition (IMT) in a 2D system was predicted \cite{[1]}.  Nevertheless, IMTs have been found in a variety of doped semiconducting 2DES (2D-electron or -hole systems) \cite{[2]}, and they still represent an open and intriguing problem for the physics of 2D systems. The pioneering work of Kravchenko et al.\ on the insulator to metal transition (IMT) in Si metaloxidesemiconductor field-effect transistors (MOSFETs) \cite{[3]} challenged the scaling theory of localization \cite{[1]}. Since then, IMTs have been found in several semiconducting 2D-electron or -hole systems (2DES) \cite{[2],[3],[4]}. In such systems, the IMT is primarily driven by doping; the system is insulating for small carrier concentrations and metallic for large carrier concentrations, with the transition defined by a critical carrier concentration. The behavior of the electrical resistivity $\rho$ as a function of temperature $T$ characterizes the insulating ($\frac{d\rho}{dT}<0$) and metallic ($\frac{d\rho}{dT}>0$) regimes. Only for a small range of concentrations near criticality in the metallic side, the system displays a non-monotonic behavior in $\rho(T)$. Most recently, Wang et al.\ reported a dimensionality-driven insulator-metal transition in Ruddlesden-Popper series \cite{[5]}.

Pristine graphene is a 2D allotrope of carbon, and possesses many unique electrical properties \cite{[6],[7],[8]}. In spite of its high crystallinity, various types of extrinsic disorder such as ad-atoms and charged impurities \cite{[9],[10],[11]}, or intrinsic ones like surface ripples, topological defects and the supporting substrate affect the electronic properties of graphene \cite{[12],[13],[14],[15],[16]}, which were also anticipated by theoretical reports \cite{[17],[18]}. Hints of IMT in graphene have been observed or predicted recently \cite{[19],[20],[21]}. Bolotin et al.\ have produced near-pristine high-mobility single layer graphene structures, in which ballistic conductance has been achieved \cite{[19]}. In their devices, near the neutrality point the resistance decreases with increasing temperature and shows the opposite behavior for higher doping levels, increasing linearly with temperature. Hwang and Das Sarma \cite{[20]} have proposed that temperature-dependent screening could explain these results.  Finally, Adam et al.\ \cite{[21]} found signatures of a percolation-driven IMT in graphene nanoribbbons. Most recently, a metal/insulator transition also has been reported in damaged graphene derivatives, such as oxidized or hydrogenated graphene \cite{[22],[23],[24],[25]}. As we shall see below, the present results on recently reported \cite{[26],[27]} atomic layers containing hybridized domains of graphene and h-BN (named as h-BNC), which have sparked a considerable amount of recent theoretical work \cite{[46],[47],[48]}, follow a completely different and novel phenomenology as compared to the above results on high-purity or damaged graphene samples where the IMT is externally imposed (by the gate) and not intrinsic and temperature dependent as in the present work.

Here, we report electrical transport measurements performed on h-BNC field effect transistors (FET), and we find that the resistance of h-BNC shows an anomalous insulator to metal transition for all measured devices with different compositions (from 40 at\% to 94 at\% carbon). With decreasing temperature, initially the resistance increases to a maximum value displaying a negative derivative $\frac{dR}{dT}$ (typical of an insulating behavior), and then, by further decreasing the temperature, the resistance drastically decreases showing a positive derivative $\frac{dR}{dT}$ typical of a metal. This behavior is not only robust under electron and hole doping by a gate potential and under an external magnetic field, but it also shows a remarkable single-parameter scaling of the $R(T)$ curves for different doping and field conditions. In particular, the doping dependence of $R(T)$ in the metallic regime is novel, anomalous, and, to some extent, counter-intuitive: In h-BNC, for all at\% carbon compositions, the metallic regime is suppressed, rather than enhanced, by doping. In the insulating regime, h-BNC displays Efros-Shklovskii variable range hopping (ES-VRH) between localized states \cite{[28],[29]}.

\section{Methodology}

\begin{figure*}
\centering
\includegraphics[width=\textwidth]{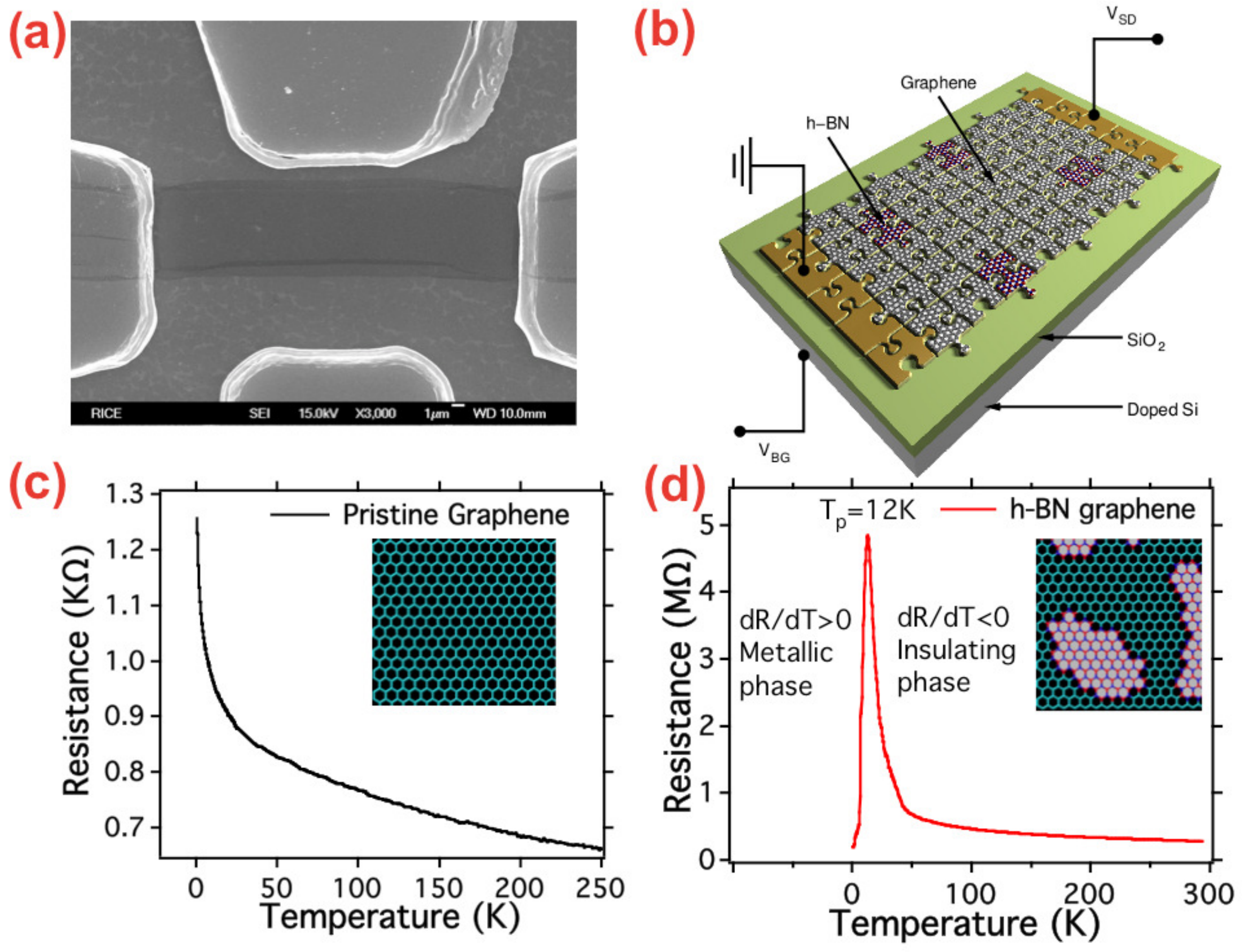}
\caption{(Color online)
Low temperature transport of h-BNC and pristine graphene FETs. (a) Scanning electron microscope image of one of our experimental devices. b, Schematic diagram of the device geometry. (c-d) Temperature dependence of the resistance of a pristine graphene (c) and a hybridized BNC with 90 at\% carbon (d) under zero magnetic field in a cooling run. The curve of h-BNC exhibits a peak at $T_\mathrm{peak} = 12$ K, signaling the insulator-metal transition. A negative derivative of the resistance ($\frac{dR}{dT}<0$) corresponds to an insulating phase, and a positive sign ($\frac{dR}{dT}>0$) is empirically associated with the metallic phase. It should be clarified that graphene was measure in a 4 terminal configuration while h-BNC was measured in 2 terminal one. The insert images of (c) and (d) are atomic models for pure graphene and hybridized h-BN graphene, respectively.
}\label{Fig1}
\end{figure*}

Hybridized h-BNC and pristine graphene samples were synthesized by thermal catalytic chemical vapor deposition. The growth details are described in Ref.~\onlinecite{[26]}. After transferring onto a heavily doped Si substrate having a 300 nm SiO$_2$ layer, these as-grown films were patterned into narrow ribbons with widths ranging from 5 to 10 $\upmu$m through optical lithography and oxygen plasma etching techniques. Subsequently, the h-BNC and pristine graphene ribbons with length 10-30 $\upmu$m were contacted with Ti/Au (3/30 nm) metal electrodes, as shown in Figure \ref{Fig1}a. The details concerning device fabrication are shown in Appendix \ref{DeviceFabrication}. Figure \ref{Fig1}b shows a schematic depicting the geometry of the device. The drain to source voltage was applied through the top two electrodes while the back gate voltage was loaded from the bottom Si$^{++}$ electrode. A High-resolution Transmission Electron Microscopy (HRTEM) image taken on the edge of the sample indicates that most of the h-BNC films are composed of three layers with a hexagonal packing model. The electron energy loss spectroscopy (EELS) performed on the film indicates that three elements (B, C and N) are uniformly distributed throughout the film with $sp^2$-hybridized bonds, as described in Appendix~\ref{SampleCharacterization}. We note that all the experimental data presented here is based on samples composed of two to three layers instead of a single layer of h-BNC. More detail of the structural characterization of h-BNC can be found in our previous work \cite{[26]}.  All the subsequent low temperature transport measurements were performed by using a variable temperature (0.5-300 K) $^3$He cryostat under magnetic fields up to 9 Tesla, and various of AC and DC set ups. The external magnetic field (H) was applied normally to the h-BNC plane, with further details provided in Appendix~\ref{MagneticFieldDependence}.

For the low temperature transport measurements, h-BNC samples were deposited on poly-silicon hetero structures, typically covered with a 300 nanometer thick insulating SiO$_2$ layer which prevents any leakage of carriers from the sample to the substrate. Samples were measured using the source-measurement unit Keithley 2602A (which has a sense input impedance $>$ 10G$\Omega$). For samples with lower resistances, we checked the obtained results by using either a Stanford Research System 830 Lock-In (input impedance $\sim$ 10 M$\Omega$) in combination with a Keithley 6221 current source (output resistance: $ > 10^{14} \Omega$ in the 2 nA to 20 nA range) or a Lakeshore 370 resistance bridge (input impedance approaching $5 \times 10^{12} \Omega$) with excitation currents ranging from 0.01 to 1 nA. We also checked that the obtained results were current independent. Samples were mounted onto male chip carriers, themselves plugged into female chip carriers adapted to rotators and wired with Polyimide insulated phosphor bronze twisted pairs where each wire was soldered to a pin whose typical separation is of nearly 1 mm and which were further insulated with heat shrinking polyolefin tubing. $^3$He was used as an exchange gas. We have measured more than five h-BNC devices with different compositions (the percentage of carbon was defined by XPS) in order to confirm this unique insulator-metal transition, and each device has also been tested several times to avoid any artifacts.

All DFT calculations have been performed using the real-space code GPAW \cite{GPAW,GPAWRev}, employing a double-zeta-polarized localized basis set, and the Perdew-Burke-Ernzerhof (PBE) exchange-correlation functional \cite{PBE}. The occupation of the Kohn-Sham orbitals was calculated at an electronic temperature of $k_BT \approx$ 0.1 eV, with all energies then extrapolated to $T$ = 0 K.  The product of the supercell dimensions and the number of k-points used in the calculations is $> 25$ \AA\ in all repeated directions. Spin polarization has been taken into account in all calculations.  Structural relaxation has been performed until a maximum force below 0.05 eV/\AA\ was reached. Transport calculations have been performed using the nonequilibrium Green's function (NEGF) method with the electronic Hamiltonian obtained from the DFT calculation.

\section{Results}

Figure \ref{Fig1}c shows the temperature dependence of the resistance for a pristine graphene sample in zero magnetic-field. To provide the resistivity instead of the raw resistance traces, one would need to know the precise number of BNC layers in the device, which in our case could not be determined. As is clearly seen, the resistance grows monotonically as the temperature decreases, which indicates that pure graphene acts as a semiconductor with zero band-gap in agreement with previous reports \cite{[6],[7],[8]}. Figure \ref{Fig1}d shows a typical temperature dependence of the resistance at zero-field for an h-BNC sample with 90 at\% carbon (the same $R$ versus $T$ curves for h-BNC devices with other at\% carbon compositions are provided in Appendix~\ref{LowTTransport}). In contrast with pristine graphene, h-BNC shows a clear insulator to metal transition with a peak at around 12 K, where a negative derivative of the resistance as a function of temperature ($\frac{dR}{dT}$) corresponds to an insulating behavior and the positive sign is empirically associated with a metallic behavior, as marked in Figure \ref{Fig1}d. Experimentally, on the insulating side of the transition, the two terminal resistance of h-BNC monotonically increases to reach a maximum (peak at $\sim$4837 k$\Omega$ as temperature decreases from room temperature to 12 K). On the metallic side of the transition, the resistance drops drastically by more than one order of magnitude ($\sim$193 k$\Omega$ as the temperature decreases from 12 K to the lowest achieved value of 500 mK). We have measured our samples in constant electrical current mode, with currents ranging from 0.01 to 1 nA, not observing any significant effect of the magnitude of the current across the transition, meaning no significant differences in the respective values of the resistivity in any range of temperatures. We remark that all the measurements are within the ohmic regime in the transition region.

\begin{figure*}
\centering
\includegraphics[width=\textwidth]{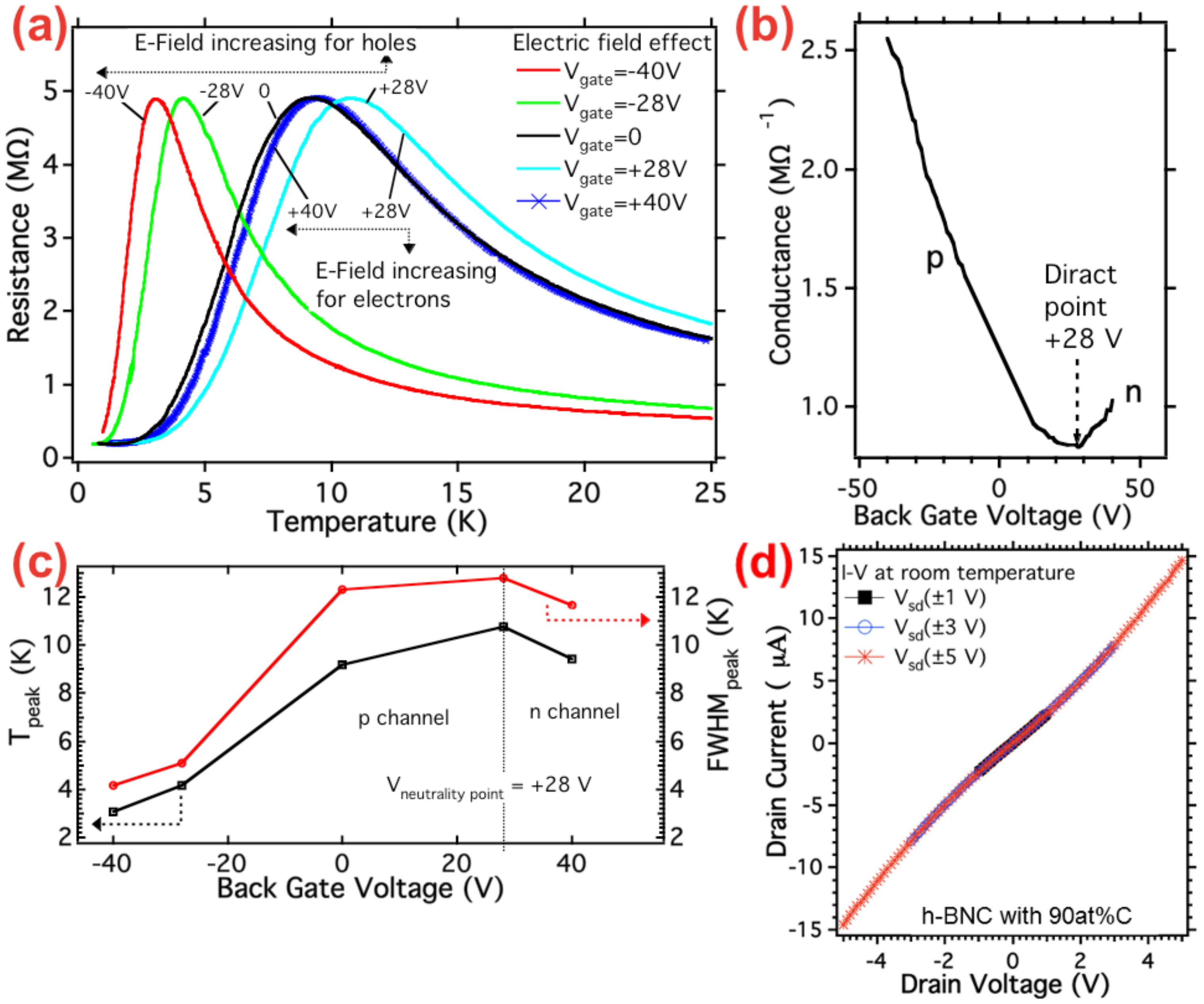}
\caption{(Color online)
Electric field dependence of the insulator-metal transition in h-BNC at 90 at\% carbon. (a) Temperature dependence of the h-BNCs resistance at gate voltage $V_\mathrm{gate}=$ +40, +28, 0, -28, -40 V in a fixed magnetic field of 3 Tesla. (b) the variation of resistance as a function of gate voltage measured at a temperature of 37.5 K. The vertical line marks the neutrality point (Dirac point) of h-BNC FET. (c) The corresponding changes of peak position and FWHM in abnormal IMT peak as a function of gate voltage in (a). (d) Room temperature I-V curves for h-BNC under different bias voltages.
}\label{Fig2}
\end{figure*}

\begin{figure*}
\centering
\includegraphics[width=\textwidth]{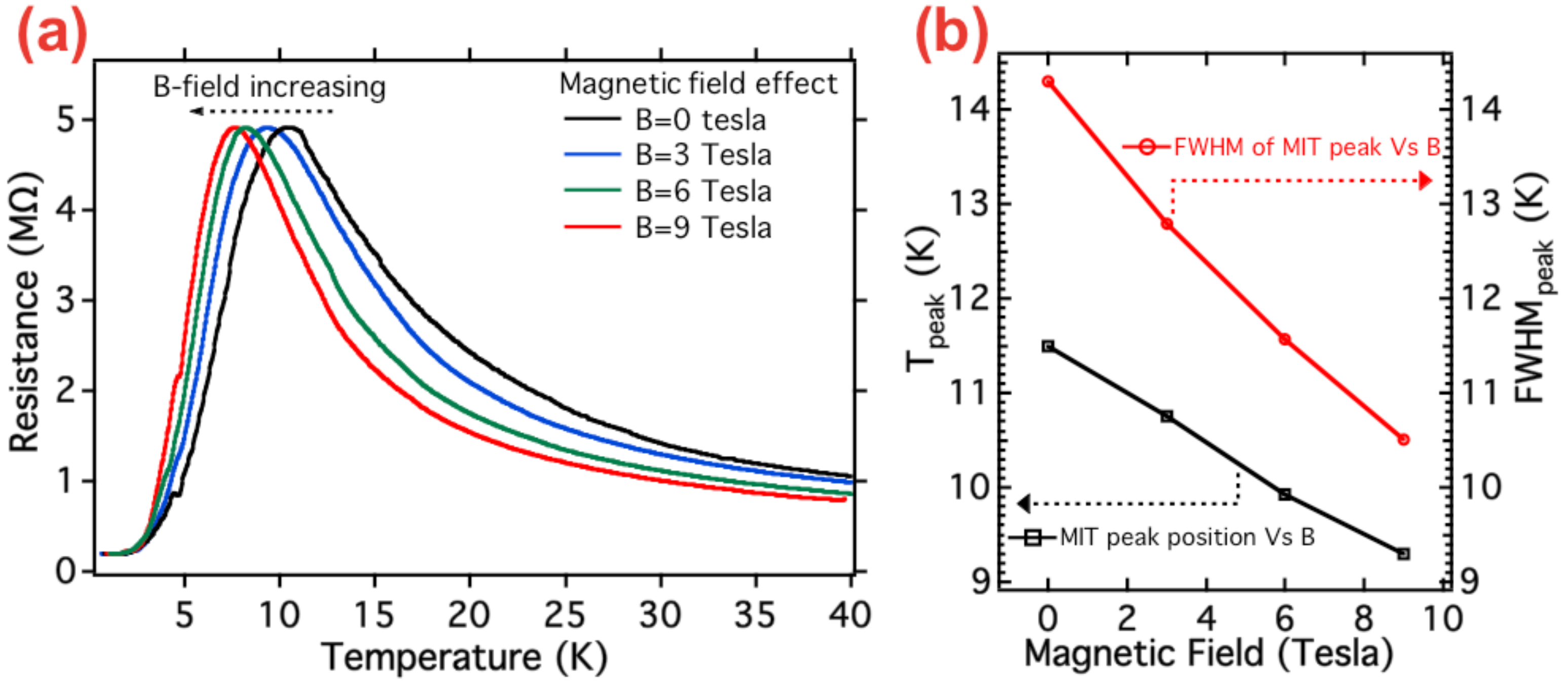}
\caption{(Color online)
Magnetic field dependence of the insulator-metal transition in h-BNC at 90 at\% carbon. (a) Temperature dependence of the h-BNCs resistance at magnetic field $B=$ 0, 3, 6 and 9 Tesla measured in zero gate voltage. The abnormal IMT point downs shift to lower temperature with the increasing magnetic field. The dashed lines are guides to the eye. (b) Changes of peak position and FWHM in the abnormal IMT peak as a function of magnetic field measured in a gate voltage of 28 V.
}\label{Fig3}
\end{figure*}

The changes in resistance as a function of back gate voltage were measured at a temperature of 37.5 K on the h-BNC FET devices. As is clearly shown in Figure \ref{Fig2}b, the resistance curves exhibit an ambipolar semiconducting behavior for h-BNC, which is similarly observed in pristine graphene. Considering the typical p-type semiconducting behavior of B-C-N mixed atomic structures as expected from both theoretical predictions and previously reported data \cite{[31],[32],[33]}, we propose that the behavior of our h-BNC films can be understood as a percolating graphene network with embedded BN domains \cite{[26],[27]}. Note that the maximum resistance value (corresponding to the Dirac point) for h-BNC is shifted to 28 V as a result of the chemical doping and boundary scattering between h-BN and graphene domains. The calculated electron-mobility of our h-BNC devices is 12 cm$^2$V$^{-1}$s$^{-1}$ with a hole-mobility of 8 cm$^2$V$^{-1}$s$^{-1}$, which is much smaller than the mobilities reported for graphene \cite{[6],[7],[8]}. This indicates that, although the $sp^2$ network is preserved in h-BNC \cite{[26],[27]}, the random distribution of h-BN clusters and associated domain boundaries are strong sources of electron scattering, typical of highly-disordered systems.

We also analyzed the IMT in h-BNC as a function of doping by measuring the $R(T)$ curves while modulating the gate voltage within the range of -40 to 40 V, as shown in Figures 2a and 2c. Figure \ref{Fig2}a shows the temperature dependence of the resistance of h-BNC at various doping levels in a fixed magnetic field of 3 Tesla. We observe in Figure \ref{Fig2}b that the conductance of h-BNC becomes a p-dominated conductance within the gate voltage range (-40 V$<V_{\mathrm{gate}}<28$ V), while it shows an n-dominated conductance when the gate voltage is larger than 28 V. Here, we found that the IMT transition temperature for h-BNC with p-dominated conductance down-shifts from 10.5 K to 3K when the gate voltage is varied from 28 V to -40 V, while the IMT transition temperature of h-BNC with n-dominated conductance drops down to 9.5 K when the gate voltage is increased from 28 V to 40 V. Meanwhile, the full width at half maximum (FWHM) of the peak observed at the IMT transition decreases as the peak position is shifted to lower temperatures. Figure \ref{Fig2}c shows the changes in position and FWHM of the IMT peak as a function of the gate voltage when measured in a magnetic field of 3 Tesla. It indicates that the IMT peak down-shifts and its FWHM decreases with increasing doping for both electrons and holes. Additional doping dependence measured under various magnetic fields is presented in the Appendix~\ref{MagneticFieldDependence}.

Figure \ref{Fig2}d reproduces a series of I-V characteristics for the h-BNC sample with 90 at\% carbon at room temperature under different bias voltages. The linear behavior seen here indicates that the contact resistance between h-BNC samples and the metallic electrodes in our experiment is not significant. Moreover, by comparing the data for the two-probe and four-probe configurations from our previous work \cite{[26]}, we found no significant differences between the two types of conductivity data obtained for our devices.

We measured the anomalous IMT in h-BNC as a function of perpendicular magnetic fields up to 9 Tesla. Figure \ref{Fig3} shows the temperature dependence of the resistance of h-BNC under several magnetic field values for zero gate voltage. As seen, the IMT transition temperature shifts to lower temperatures when an external magnetic field is applied. The higher the magnetic field the larger the observed down-shift. Notice also that the peak at the IMT becomes narrower and sharper as the peak position is down-shifted to lower temperatures. The FWHM of the IMT peak decreases with increasing magnetic field. Figure \ref{Fig3}b indicates the changes in position and FWHM of the IMT peak as a function of magnetic field measured under a gate voltage of 28 V. As clearly seen by the position of the peak, the FWHM drops monotonically with increasing magnetic field. The dependence of IMT on magnetic field measured for various gate voltages is listed in Appendix \ref{MagneticFieldDependence}. These observations on the IMT and its dependence on gate voltage and magnetic field are reproducible and qualitatively similar for different h-BNC samples having different carbon content. We remark that in contrast to 2DEG (MOSFETS) \cite{[4]} where the B field kills the IMT transition, in the present  case for h-BNC the IMT transition shifts to lower temperatures but remains robust.

\section{Discussion and Conclusions}

\begin{figure*}
\centering
\includegraphics[width=\textwidth]{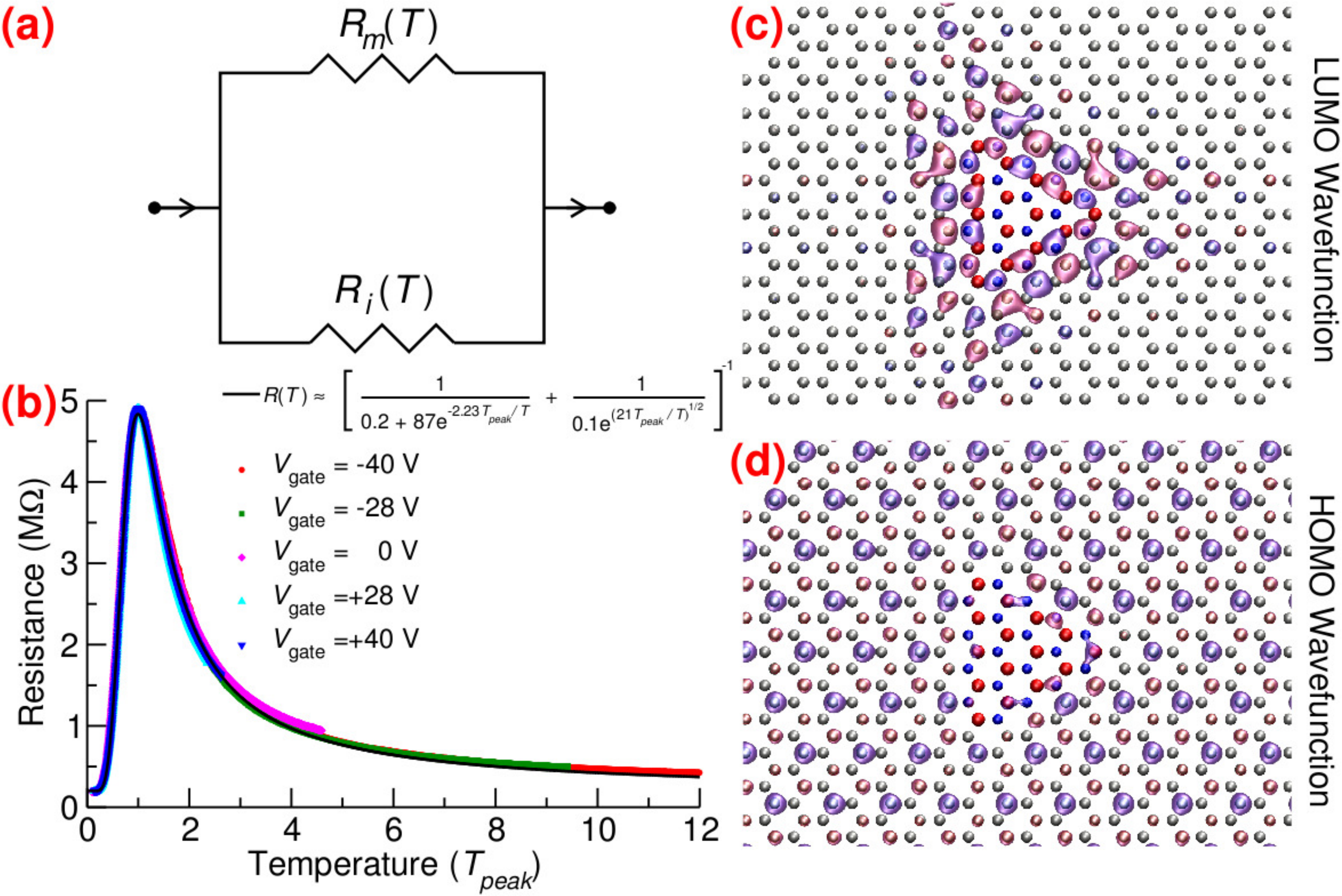}
\caption{(Color online)
Theoretical model for the insulator-metal transition in h-BNC: (a)                  Phenomenological parallel-resistor model for the IMT. (b) Single-parameter scaling leading to the collapse of all five $R(T)$ curves shown in Figure \ref{Fig2}a, for different doping levels, into a single universal curve. The parallel-resistor model (black line) is compared to the experimental resistance for the five different gate voltages. Schematics of a N terminated zigzag edged nanodomain in h-BNC with wavefunctions depicted as isosurfaces of $\pm0.025 $ \AA$^{-3/2}$ for (c) localized edge states at the nanodomain boundaries and (d) extended graphene-like states. B, N, and C atoms are colored red, blue and grey, respectively.
}\label{Fig4}
\end{figure*}

Figure \ref{Fig2}a shows that the peak and minimum values for the resistance as a function of temperature are remarkably independent of the doping levels for the 90 at\% carbon sample (for other compositions see Appendix~\ref{LowTTransport}). This suggests that doping and temperature effects must be described in a combined way and that the $R(T)$ curves should display a single-parameter scaling. In other words, all resistance curves should follow a universal function of the rescaled temperature $T/T_\mathrm{peak}$. Single-parameter scaling is a signature feature of IMT in semiconducting 2DES. Figure \ref{Fig4}b shows that indeed all $R(T)$ curves for gate voltages -40V, -28V, 0V, +28V and +40V collapse into a single universal curve when plotted as a function of the rescaled temperature. The collapse is nearly perfect, which is an indication of the similarity between the IMT in h-BCN and in 2DES. Remarkably, the scaling works equally well in both the metallic and the insulating regions.

However, there are important and surprising differences between the IMT in h-BNC and 2DES. First, as we mentioned, the IMT in h-BNC is ``temperature-driven'' while in 2DES the transition is ``density-driven''. By this we mean that, in 2DES, the occurrence of metallic or insulating regimes is primarily determined by the carrier density and it is fairly independent of the temperature, so the transition is defined by a critical density. On the other hand, in h-BNC, both metallic and insulating regimes can be observed for all carrier densities studied in our work, and the transition is characterized by a critical temperature separating the two regimes. A second important difference is that, contrary to 2DES, in h-BNC doping with either electrons or holes seems to suppress the metallic state, i.e., the range of temperatures in which the system is metallic decreases with doping. This is a novel and anomalous behavior.

The robust non-monotonic behavior of $R(T)$ for all doping levels studied in h-BNC led us to picture this system as described by the coexistence of metallic and insulating conducting channels, as described in models for the IMT such as in manganites \cite{[34],[35],[36]}. This picture may also be adequate for a composite system like h-BNC. If we assume these mechanisms to act independently, we can formulate a phenomenological parallel-resistor model to describe the $R(T)$ curves (Figure \ref{Fig4}a): $R(T)=[1/R_m(T)+1/R_i(T)]^{-1}$, where $R_m(T)$ and $R_i(T)$ are the metallic ($\frac{dR_m}{dT} >$ 0) and insulating ($\frac{dRi}{dT} <$ 0) components of the resistance, respectively. In the metallic regime, $T\ll T_{\mathrm{peak}}$, then $R(T)\sim R_m(T)$. We find that $R_m(T)$ has the empirical exponential form also typical of 2DES \cite{[37]}
\begin{eqnarray}
R_m(T) = R_0+R_1 \exp(-T_m/T),
\end{eqnarray}
for which no final theory exists as yet \cite{[5]}, with different proposals as diverse as temperature-dependent screening \cite{[38]}, percolation \cite{[39]}, charging of traps \cite{[40]}, and an analog of the Pomeranchuk effect \cite{[41]}. In the insulating regime, $T\gg T_{\mathrm{peak}}$, $R(T)\sim R_i(T)$, the conductance in 2DES that is often associated \cite{[42],[43]} with either Mott \cite{[44]} or Efros-Shklovskii \cite{[28],[29]} (ES) variable-range hopping (VRH) between localized states. Our data is better described by the ES-VRH model, also known as VRH under a Coulomb gap:
\begin{eqnarray}
R_i(T) = R_2 \exp((T_i/T)^{1/2}).
\end{eqnarray}
Figure \ref{Fig4}b shows a perfect agreement between this parallel-resistor model and the experimental data. The best-fit parameters for the 90 at\% carbon sample are $R_0 = 0.2$ M$\Omega$, $R_1 = 87$ M$\Omega$, $R_2 = 0.1$ M$\Omega$, $T_m = 2.23T_{\mathrm{peak}}$ and $T_i = 21T_{\mathrm{peak}}$. The doping (or gate potential) dependence of resistance is automatically built-in in the proportionality between $T_i$, $T_m$, and $T_{\mathrm{peak}}$, which is a direct consequence of the single-parameter scaling of the data.   The low temperature limit of the resistance is independent of doping, with $R(T) \sim R_0 \sim 0.2$ M$\Omega$. This we attribute to scattering off the insulating BN nanodomains, since it agrees qualitatively with the resistances obtained from non-equilibrium Green's function (NEGF) calculations  of the transmission function through several model h-BNC systems.  Specifically, we performed density functional theory (DFT) calculations \cite{GPAW,GPAWRev} for various widths of zigzag terminated h-BN nanoribbbons and triangular N-terminated zigzag edged nanodomains, both embedded in graphene. The details on both fitting procedures and calculations are shown in Appendix \ref{model} and \ref{theory}.

The model described above implies the coexistence of percolating graphene-like electronic states that are responsible for the metallic channel and a distribution of localized electronic states through which VRH occurs. These states must coexist at the Fermi level for all considered doping levels, but must also be spatially separated, so that the two channels are uncoupled. From DFT calculations for both N-terminated zigzag edged nanodomains and zigzag edged nanoribbons embedded in graphene, we find indeed a coexistence of localized edge states at the nanodomain boundaries (Figure \ref{Fig4}c) and extended graphene-like states (Figure \ref{Fig4}d) near the Fermi level (separated by about 0.2 eV and 0 eV respectively), for the range of doping considered here. Since these embedded extended BN ribbon and small triangular BN domain calculations describe the limiting cases of large and small h-BN nanodomains, it is reasonable to assume that both localized edge states and extended graphene-like states are present in the experimental system. Disorder and potential fluctuations will likely broaden the energy distribution of localized edge states, with Coulomb repulsion causing the insulator-like ES-VRH conduction as opposed to metal-like band conduction. Spatial separation of the two channels will occur by the simultaneous percolation of ``metallic regions'' and ``insulating regions''.

Finally, we comment on the doping dependence of the characteristic insulating and metallic temperatures, $T_i$ and $T_m$, which are both proportional to $T_{\mathrm{peak}}$. Since $T_{\mathrm{peak}}$ decreases upon doping with either electrons or holes, so do $T_i$ and $T_m$.  This suggests that doping affects the resistance of h-BNC via a common mechanism for both metallic and insulating regimes. As described in the Appendix~\ref{model}, the decrease of $T_i$ with doping may be qualitatively understood within the well-established theory of ES-VRH, since $T_i$ should decrease with increasing screening, which should by itself increase with doping. However, the decrease of $T_m$ with doping does not fit the phenomenology of the metallic phase of semiconducting 2DES nor any of the related models. In fact, it shows precisely the opposite behavior. Of course, graphene systems are in many ways different than semiconductor 2DES, not only for their band dispersions and DOS profile near the Fermi level, but also because in doped 2DES carriers are either electrons or holes, but in h-BNC (as in graphene), potential fluctuations cause the coexistence of electron and hole puddles \cite{[25]}. Therefore, the present results represent a novel type of IMT in 2D electronic systems, and they will likely open a new pathway in the investigation of such systems.


\begin{acknowledgments}

P.M.A.\ acknowledges support from the Office of Naval Research (ONR) through MURI program, the Basic Energy Sciences division of the Department of Energy (DOE) and Rice University startup funds. L.S.\ is supported by DOE-BES program DE-SC0001479 for device fabrication and measurement. L.B.\ is supported by DOE-BES through award DE-SC0002613 and by the NHMFL-UCGP program for measurement. The NHMFL is supported by NSF through NSF-DMR-0084173 and the State of Florida. D.J.M., S.K.\ and A.R.\ acknowledge financial support from the Spanish ``Juan de la Cierva'' program (JCI-2010-08156), Spanish MICINN  (FIS2010-21282-C02-01), ACI-Promociona (ACI2009-1036), ``Grupos Consolidados UPV/EHU del Gobierno Vasco'' (IT-319-07), the European Union through the FP7 e-I3 ETSF (Contract Number 211956), and THEMA (Contract Number 228539) projects and the Ikerbasque Foundation. R.B.C.\ acknowledges financial support from Brazilian agencies CNPq, CAPES, FAPERJ and INCT --- Nanomateriais de Carbono and the ONR MURI program. S.G.L.\ acknowledges supported from National Science Foundation Grant No.\ DMR07-05941 and the ONR MURI program.  L.C.\ is supported by ONR MURI program (Award No N00014-09-1066) for sample growth. K.S.\ acknowledges support from the National Science Foundation Major Research Instrumentation program (NSF-MRI) DMR-0619801 and the Department of Energy National Nuclear Security Administration (DOE-NNSA) DE-FG52-05NA27036 and the PVAMU Title III program US department of Education. We thank Prof.\ Feng Liu (University of Utah), Prof.\ Anchal Srivastava (Banaras Hindu University) and Dr.\ Zheng Liu for useful discussions.
\end{acknowledgments}

\appendix

\section{Device fabrication}\label{DeviceFabrication}

\begin{figure}
\centering
\includegraphics[width=\columnwidth]{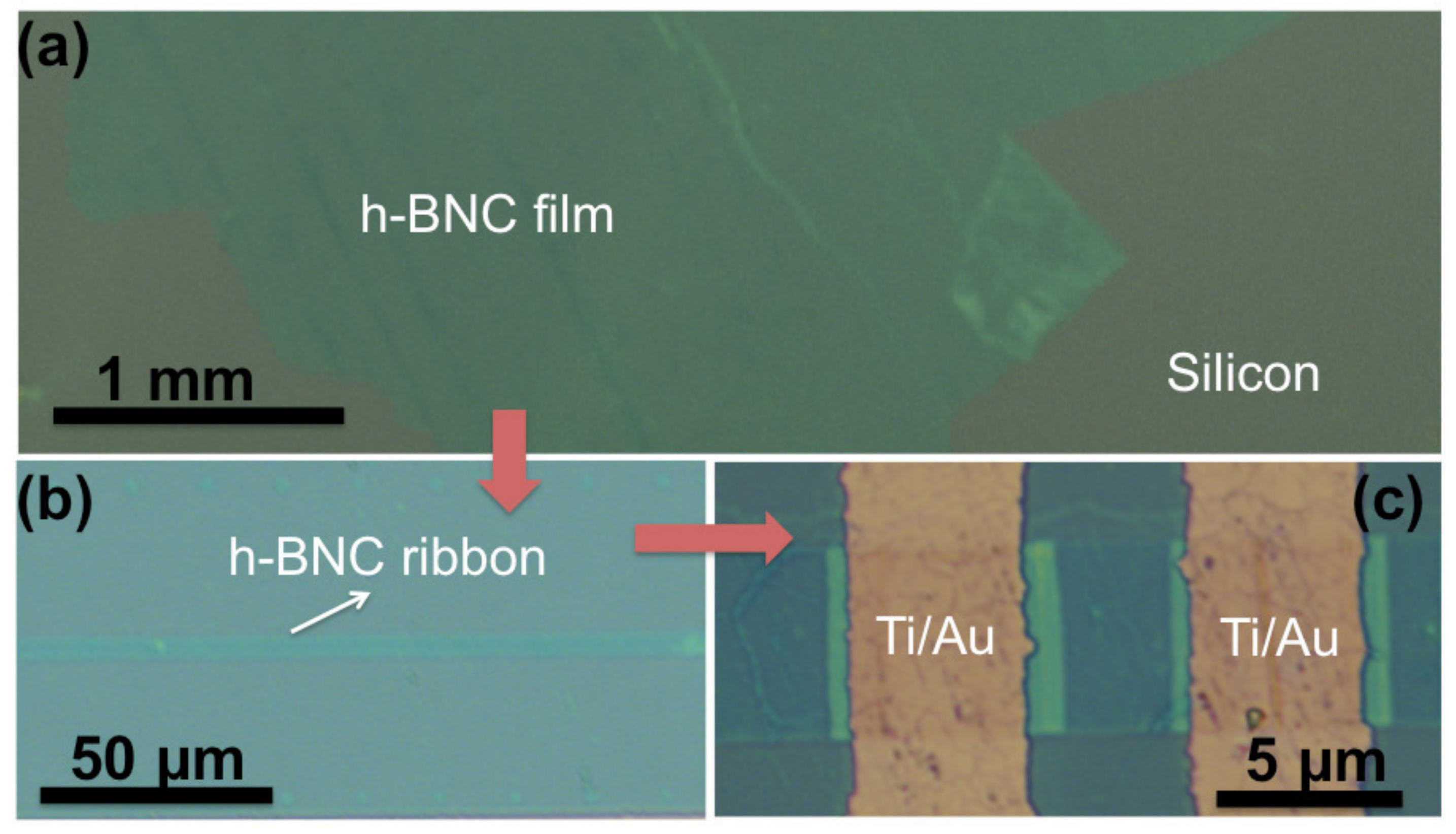}
\caption{Optical microscopy images of a transferred h-BNC film (a) and ribbon (b) on a Silicon substrate. The patterned ribbon was fabricated by using optical lithography and oxygen plasma etching. (c) Optical image of a typical h-BNC ribbon electrically contacted with Ti/Au pads of thickness ranging from 3 to 30 nm.}\label{FigS1}
\end{figure}

Ultrathin h-BNC films were directly grown on copper foil. After growth, PMMA (Poly Methyl Methacrylate) was spin-coated on the Cu substrates containing grown h-BNC. Subsequently, the sample was immersed into a diluted nitric acid solution to etch away the underlying Cu substrate. The h-BNC films supported by the PMMA remained floating on the liquid solution. The floating h-BNC films were thus transferred onto desired substrates, while PMMA was removed by using acetone and isopropanol solvents.

To realize the ribbon-geometry, ultrathin h-BNC films were coated with the S1813 photo resist. A UV-mask aligner was used to produce the 5-10 $\upmu$m linear-patterns on the coated films. After developing, the sample was placed into an oxygen plasma chamber to etch away the residual parts, leaving only the ribbons, as shown in Figure \ref{FigS1}a. After removing the photo resist with acetone, a Ti/Au metallization was used to create Ohmic contacts on the h-BNC ribbon by optical lithography and E-beam evaporation, as shown in Figure \ref{FigS1}b.

\section{Sample characterization}\label{SampleCharacterization}

\begin{figure}
\centering
\includegraphics[width=\columnwidth]{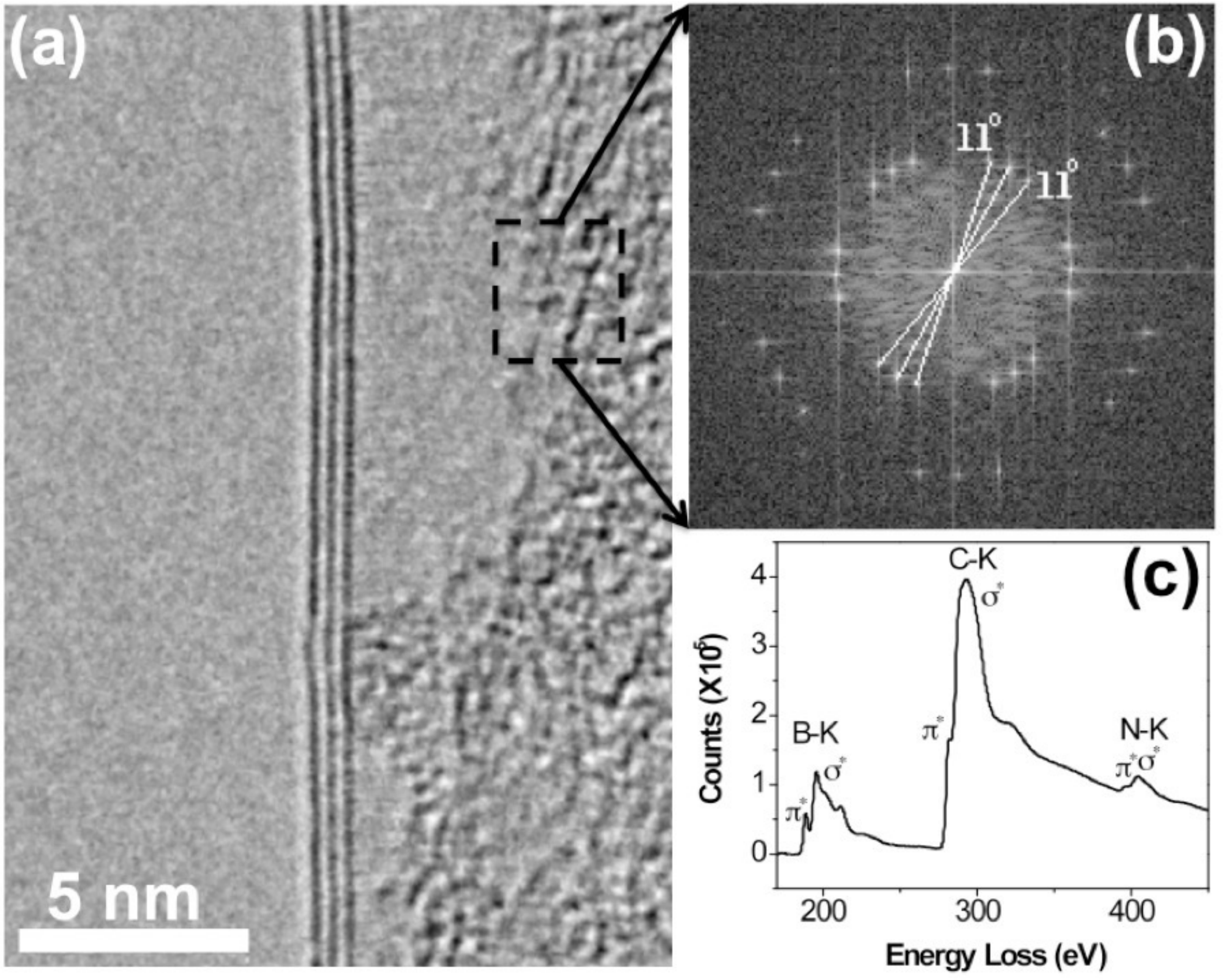}
\caption{(a) High-resolution Transmission Electron Microscopy (HRTEM) image taken on the edge of the sample indicates most of our tested films are composed of three layers. (b) The fast Fourier transform reveals a three-layer packing region with rotational angle of 11$^\circ$ in our sample.  (c) The core Electron Energy Loss Spectra (EELS) clearly shows B, C and N are uniformly distributed throughout the film.}\label{FigS2}
\end{figure}

Figure \ref{FigS2}a shows a typical high resolution TEM image of the edge of our tested films. It can be clearly seen that our h-BNC films mostly consist of three layers. The Fast Fourier transform (FFT) pattern in Figure \ref{FigS2}b taken from the plane of the films shows three sets of hexagonal spots with a rotational angle of 11$^\circ$, which indicates that our h-BNC is a three-layer hexagonal packing region. The electron energy loss spectroscopy (EELS) performed on the h-BNC film to determine their chemical composition and structure is shown in Figure \ref{FigS2}c. This proves that the films consist of B, C and N. Those three elements are sp$^2$ hybridized and uniformly distributed throughout the film. More details of the structure characterization of h-BNC can be found in our previous work \cite{[26]}.

\section{Low temperature transport measurement for h-BNC}\label{LowTTransport}

\begin{figure*}
\centering
\includegraphics[width=0.8\textwidth]{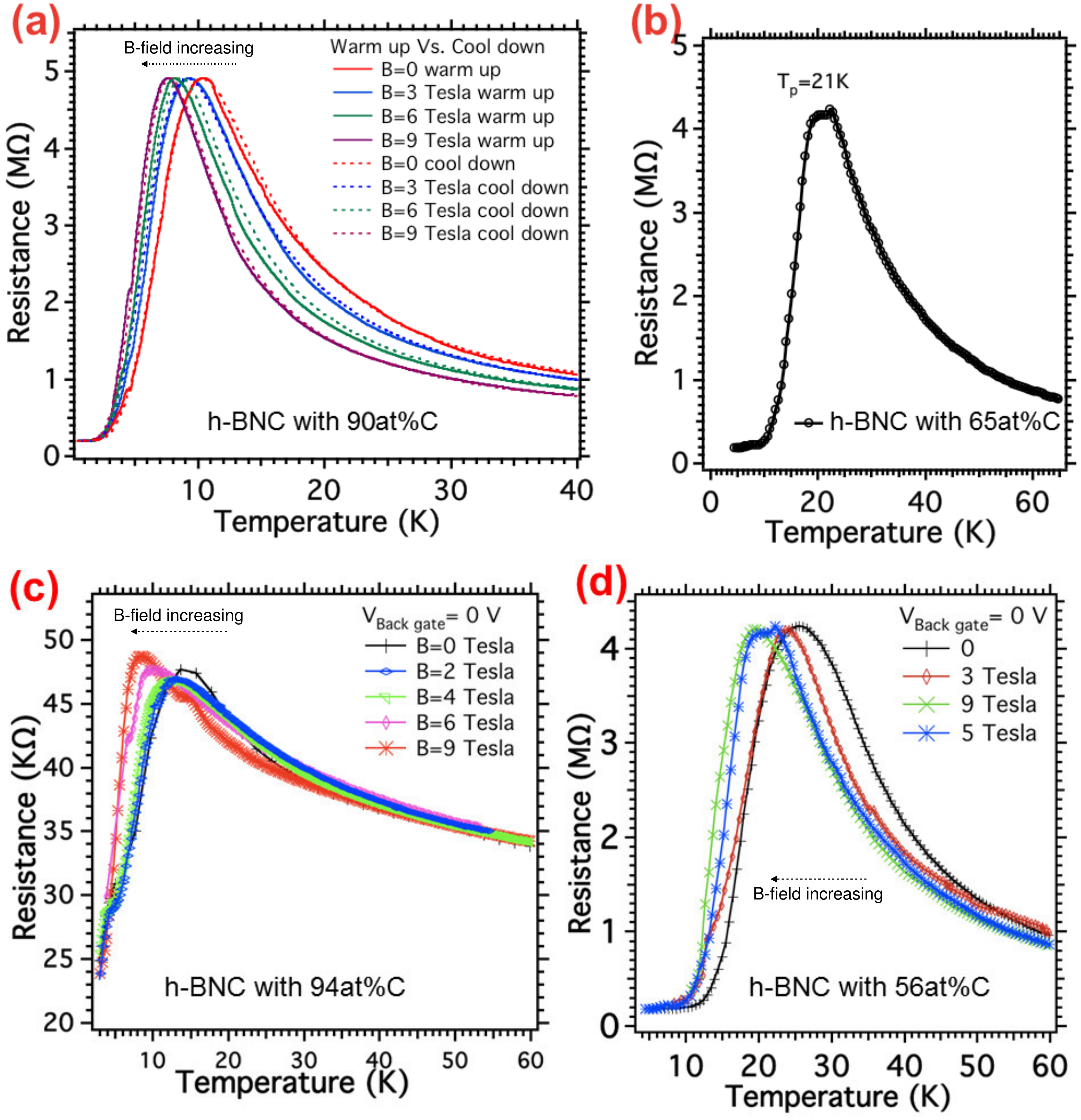}
\caption{Temperature dependence of the resistance for h-BNC films with (a) 90 at\% carbon under zero gate voltage and under various magnetic fields measured either in warming up or cooling down sweeps. Insulator to metal transition observed in different h-BNC devices with (b) 65 at\%, (c) 94 at\%, and (d) 56 at\% carbon.
}\label{FigS3}
\end{figure*}

\begin{table}
\centering
\includegraphics[width=\columnwidth]{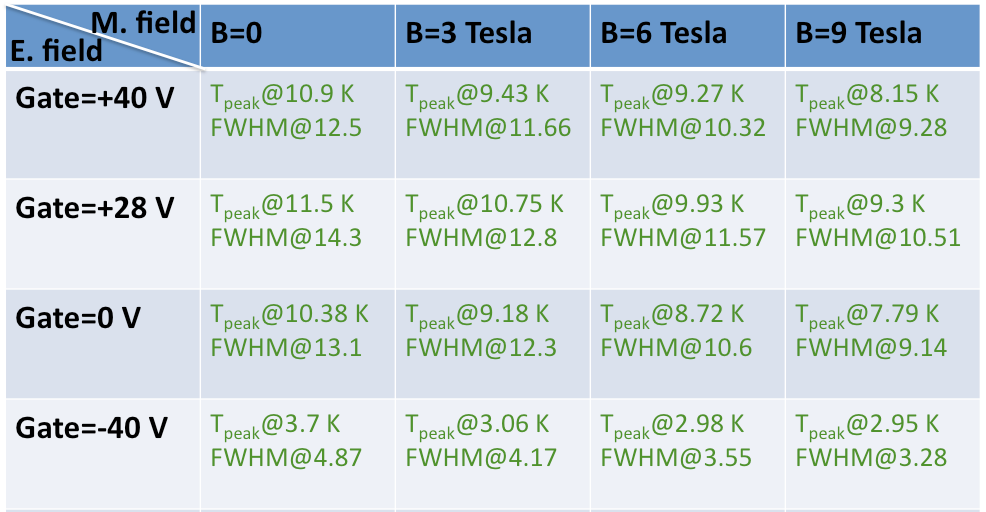}
\caption{Comparison of insulator-metal transition peak of the h-BNC with 90 at\% carbon at various magnetic fields and back gate voltages}\label{Table1}
\end{table}

\begin{figure}
\centering
\includegraphics[width=0.9\columnwidth]{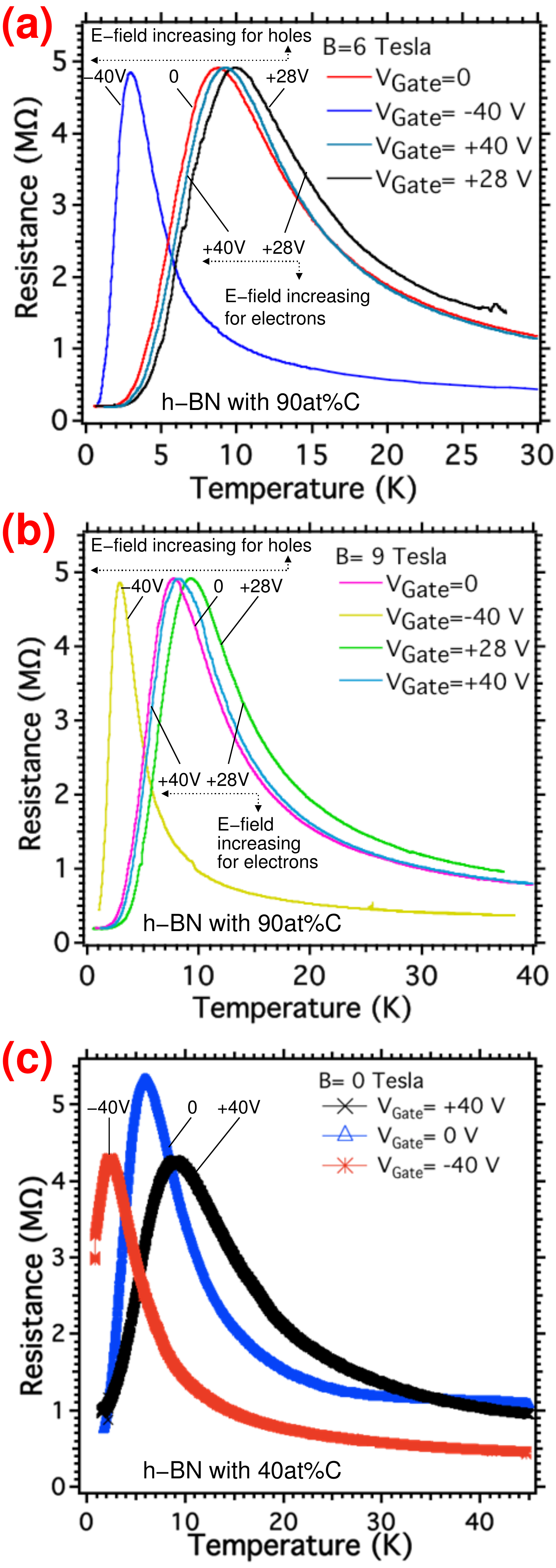}
\caption{Temperature dependence of the resistance of a h-BNC film under gate voltages of  $V_{\mathrm{gate}}$ =+40, +28, 0, -40 V, and under fixed magnetic field of (a) 6 Tesla and (b) 9 Tesla for a 90 at\% carbon device, and (c) 0 Tesla for a 40 at\% carbon device.
}\label{FigS4}
\end{figure}

Figure \ref{FigS3}a shows a typical resistivity measurement under zero gate electric field and under magnetic fields up to 9 Tesla for the h-BNC sample with 90 at\% carbon. It indicates from the warming and cooling down temperature sweeps that a very small hysteresis is observed between temperature up and down sweeps. The $R(T)$ curves under different magnetic fields also follow the single-parameter scaling by $T_{\mathrm{peak}}$ as discussed in the main text

Figures \ref{FigS3}b-d and \ref{FigS4}c show several typical traces of the resistance as a function of temperature for h-BNC films with different carbon aspect ratios, including 65 at\%, 94 at\%, 56 at\%, and 40 at\% carbon percentages, respectively. We have observed the transition in all of our samples except for the pure material (C or BN). These devices were in fact measured at two independent laboratories with different experimental set-ups by independent groups,
and in both circumstances the same behavior, namely an insulator to metal transition, was observed in our samples. Notice, how even samples displaying much lower resistances (nearly 2 orders of magnitude smaller) such as the device fabricated from BNC containing 94 at\% carbon percentage also display the same transition (Fig. 7 (c)). This, by itself, is enough to discard any instrumental artifact, such as leakage currents or the proximity of the sample's resistance to the input impedance of the measuring device.

\section{Electric field dependence of IMT in h-BNC under various magnetic fields}

Figures \ref{FigS4}a, b, and c show the temperature dependence of the resistance of a h-BNC film for various applied back gate voltages under fixed magnetic fields of 6 and 9 Tesla for a 90 at\% carbon device, and 0 Tesla for a 40 at\% carbon device, respectively. It is clearly seen that the peak position defining the IMT moves to lower temperatures with increasing back gate voltage for either electrons (within the range of -40 V to 28 V) or holes (within the range of +28 V to +40 V). This peak becomes sharper as it moves to lower temperatures.

Figure \ref{FigS4}c shows that the peak resistance for the 40 at\% sample is doping-dependent, which implies in the absence of single-parameter scaling, even though the IMT is very clear in this sample. This may be an indication that IMT for samples with low carbon content may follow a slightly different physics than the IMT for samples with high carbon content. Indeed, it seems reasonable that a percolating graphene network with embedded ``insulating'' BN clusters should have different electrical transport properties than a percolating BN network embedded  with ``metallic'' graphene clusters. Further experimental and theoretical work is needed do address this issue.

\section{Magnetic field dependence of IMT in h-BNC under various gate electric fields}\label{MagneticFieldDependence}

\begin{figure}
\centering
\includegraphics[width=0.9\columnwidth]{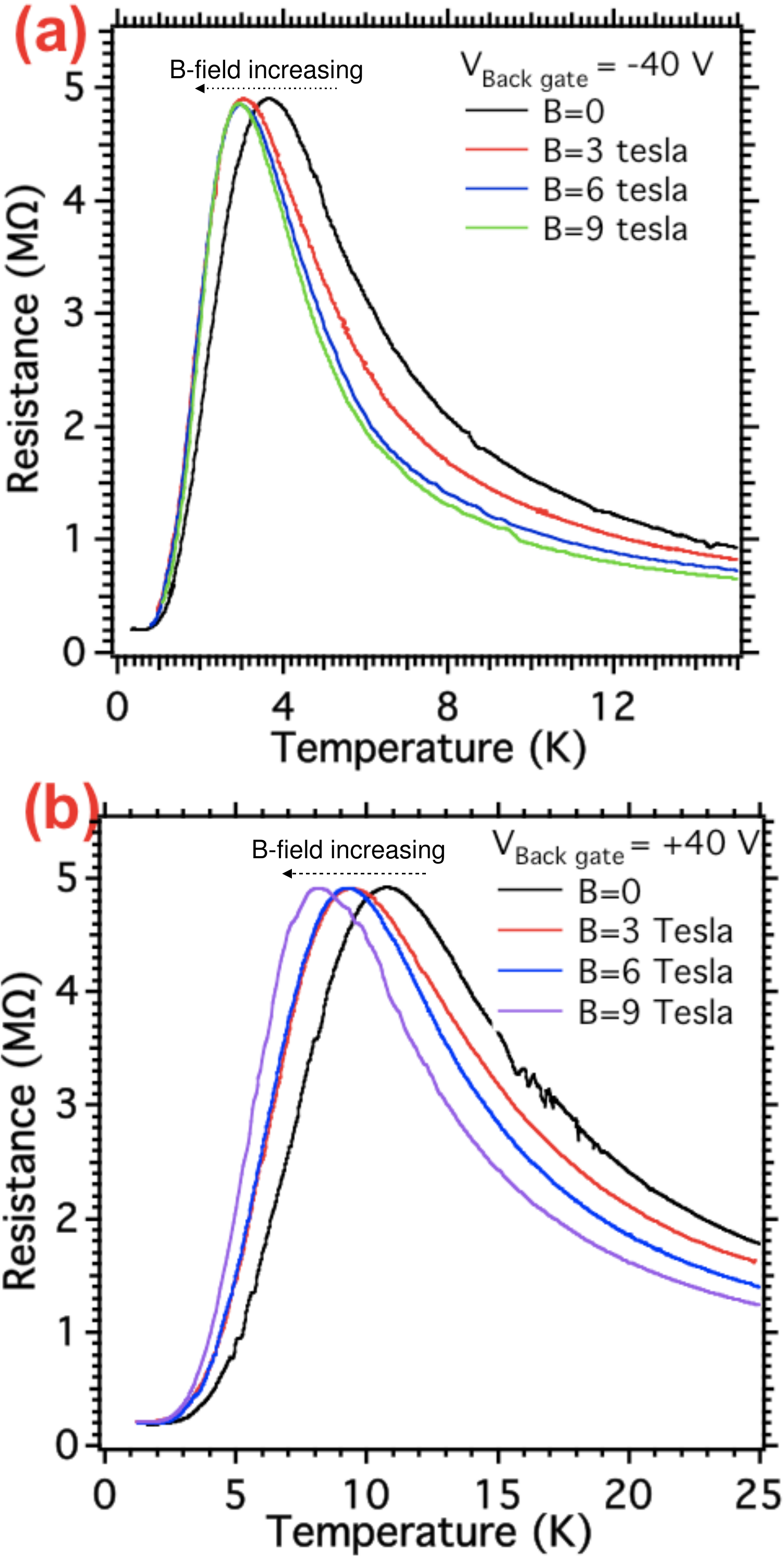}
\caption{Temperature dependence of the resistance of a h-BNC film under magnetic fields of $B$ = 0, 3, 6 and 9 Tesla, measured for back gate voltage values of (a) -40 V and (b) +40 V for a 90 at\% carbon device.
}\label{FigS5}
\end{figure}

Figure \ref{FigS5}a and \ref{FigS5}b show the IMT behavior of h-BNC as a function of magnetic field for fixed values of back gate voltages ranging from -40 V to +40 V, respectively. It indicates that the IMT transition peak shifts to lower temperatures and becomes narrower when an external magnetic field is applied. The detailed dependence of the IMT observed in h-BNC on either magnetic or electric field is shown in Table \ref{Table1}. The lowest observed transition temperature is 2.95 K under an applied field of 9 Tesla and with an applied back gate voltage of -40 V on the h-BNC film.

\section{Data analysis and modeling}\label{model}

\begin{figure}
\centering
\includegraphics[width=\columnwidth]{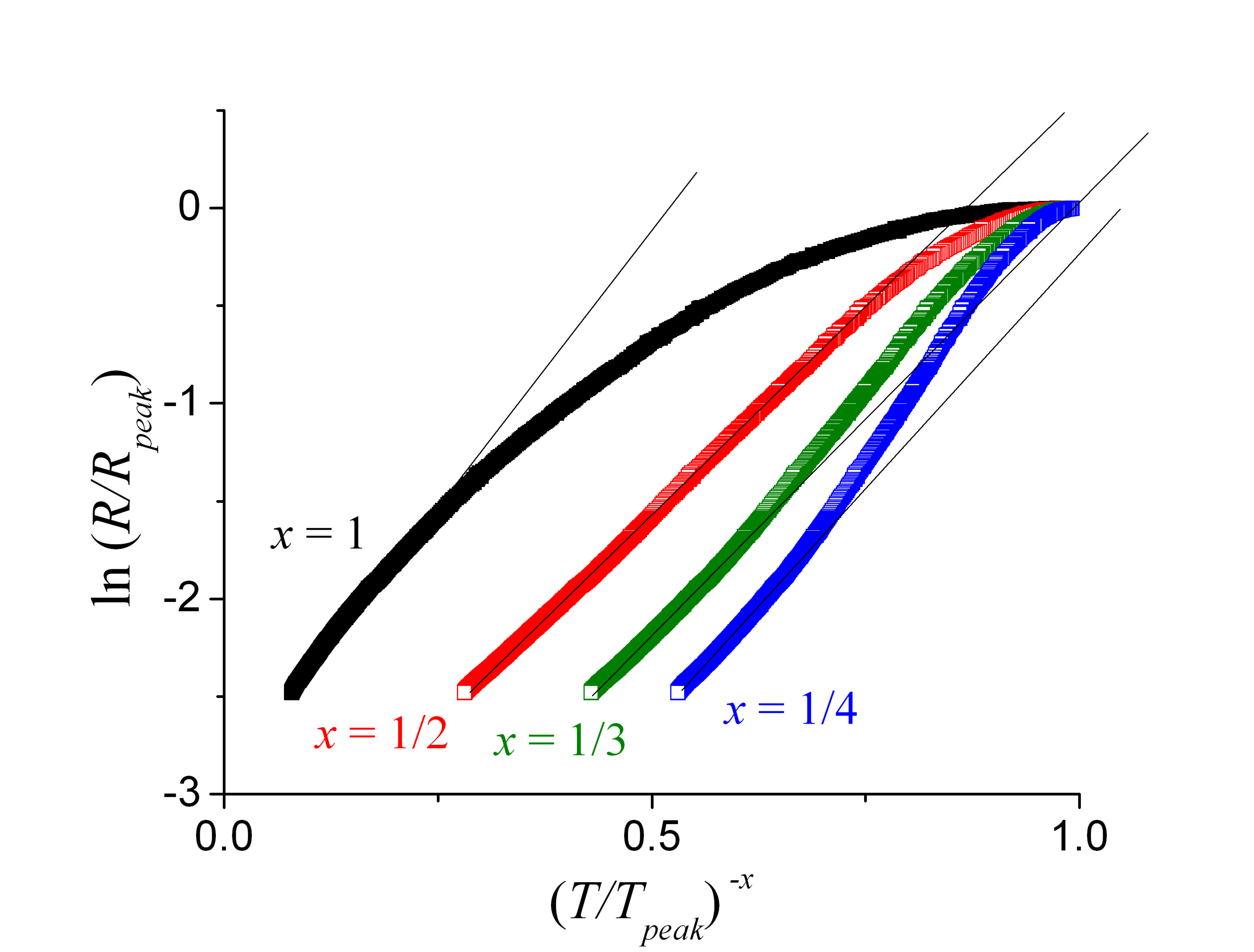}
\caption{Data fitting of the insulating portion of the $R(T)$ data. Black, red, green and blue curves correspond to $x$ values of 1, 1/2, 1/3 and 1/4, respectively. The ES-VRH mechanism ($x$ = 1/2) gives the best fit of our data to a straight line}\label{FigS6}
\end{figure}

\begin{figure}
\centering
\includegraphics[width=\columnwidth]{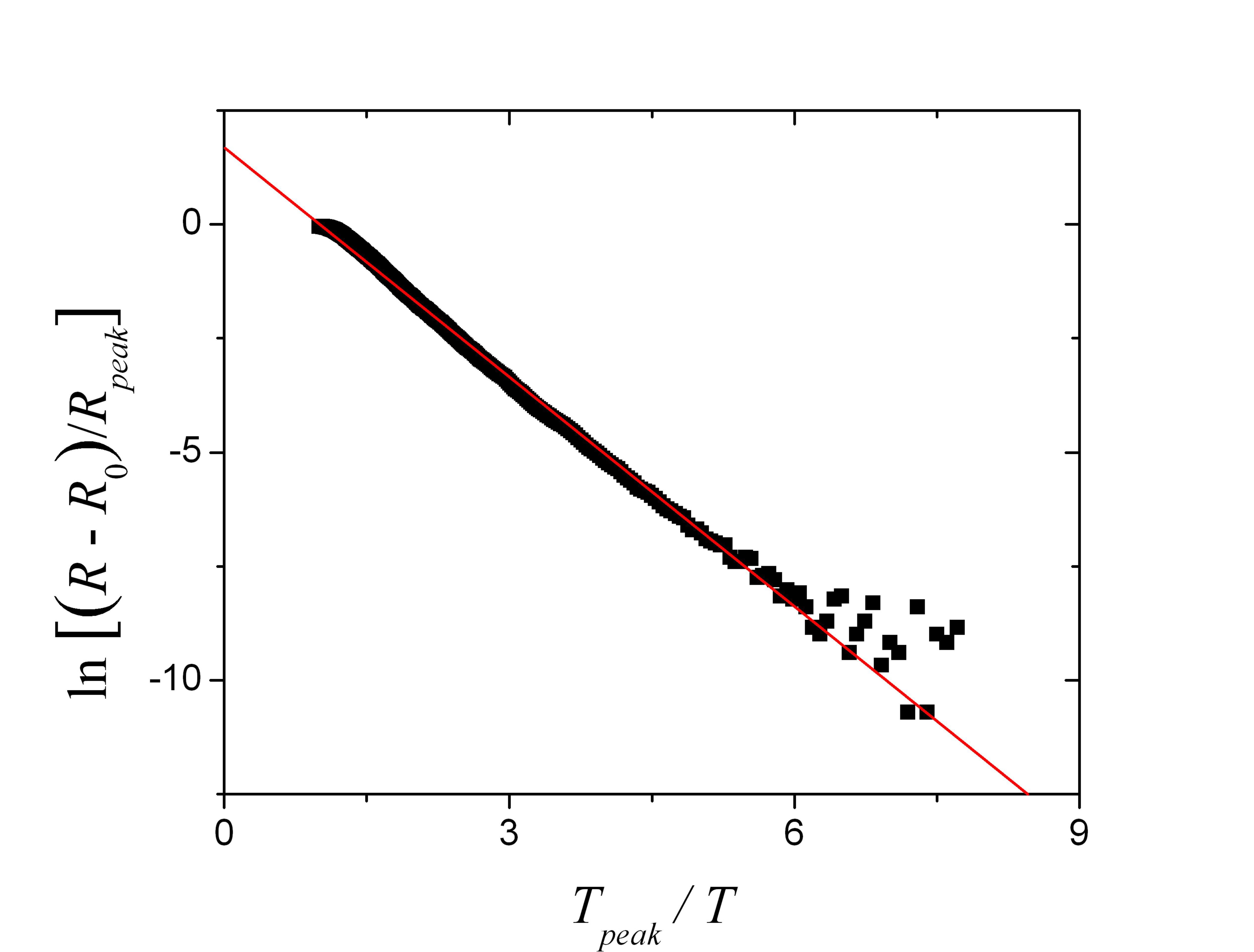}
\caption{Data fitting of the metallic portion of the $R(T)$ data.}\label{FigS7}
\end{figure}

Here we present details of our data analysis which allows for a clear identification of the analytical expressions associated with metallic- and insulating-type electronic conduction. We start by analyzing the insulating behavior. Figure \ref{FigS6} shows the $R(T)$ data for -40~V gate voltage in the insulating region ($T > T_{\mathrm{peak}}$) at zero magnetic field. In the figure, $\ln(R/R_{\mathrm{peak}})$ is plotted against $(T/T_{\mathrm{peak}})^{-x}$, so if the data follows a straight line it is an indication that the empirical law $R(T) \sim \exp[(T/T_i)^{-x}]$ is obeyed ($T_i$ being proportional to $T_{\mathrm{peak}}$). Different values of $x$ correspond to analytical expressions commonly associated to different physical mechanisms for conduction in insulating materials. For instance, $x$ = 1 corresponds to Arrhenius activation, which would imply the existence of a gap; $x$ = \nicefrac{1}{2} is the Efros-Shklovskii variable-range hopping (for any dimensionality); and $x$ = \nicefrac{1}{3} and $x$ = \nicefrac{1}{4} correspond to Mott variable-range hopping for two and three dimensions, respectively. As indicated by the straight lines (guides to the eye), the Efros-Shklovskii hopping is clearly favored in a larger temperature range with respect to the other fitting forms.

Efros-Shklovskii variable-range hopping (ES-VRH), also known as VRH under a Coulomb gap is a well-known conduction mechanism in insulators \cite{[28],[29]} and it requires the following ingredients: (1) a distribution of localized states around the Fermi level and (2) long-range Coulomb repulsion between electrons in these localized states. These conditions will lead to a Coulomb gap in the electronic DOS that in two dimensions is actually a ``soft gap'', i.e. a linear dip in the DOS precisely at the Fermi level. VRH between states at such non-constant DOS leads to the ES-VRH expression for the resistivity $R(T) \sim \exp [(T/T_i)^{-\nicefrac{1}{2}}]$, with
\begin{eqnarray}
T_i = \beta \frac{e^2}{\kappa a}
\end{eqnarray}
where $\beta$ is a constant, $a$ is the localization radius of the states and $\kappa$ is an effective dielectric constant. The doping dependence of $T_i$, that decreases for both electron or hole doping, is consistent with a larger screening of the Coulomb interactions upon enhancing the number of carriers in the percolating graphene network around the BN islands.

We now turn our analysis to the metallic behavior. Figure \ref{FigS7} shows the $R(T)$ data for +28 V gate voltage in the metallic region ($T < T_{\mathrm{peak}}$) at zero magnetic field. In the figure, $\ln[(R-R_0)/R_{\mathrm{peak}}]$ is plotted against $T_{\mathrm{peak}}/T$. The data follow very closely a straight line, indicating that the empirical law $R(T) = R_0 + R_1\exp(-T_m/T)$ is obeyed ($T_m$ also being proportional to $T_{\mathrm{peak}}$).

\section{Theoretical calculations for model h-BNC systems}\label{theory}

\begin{figure}
\centering
\includegraphics[width=0.9\columnwidth]{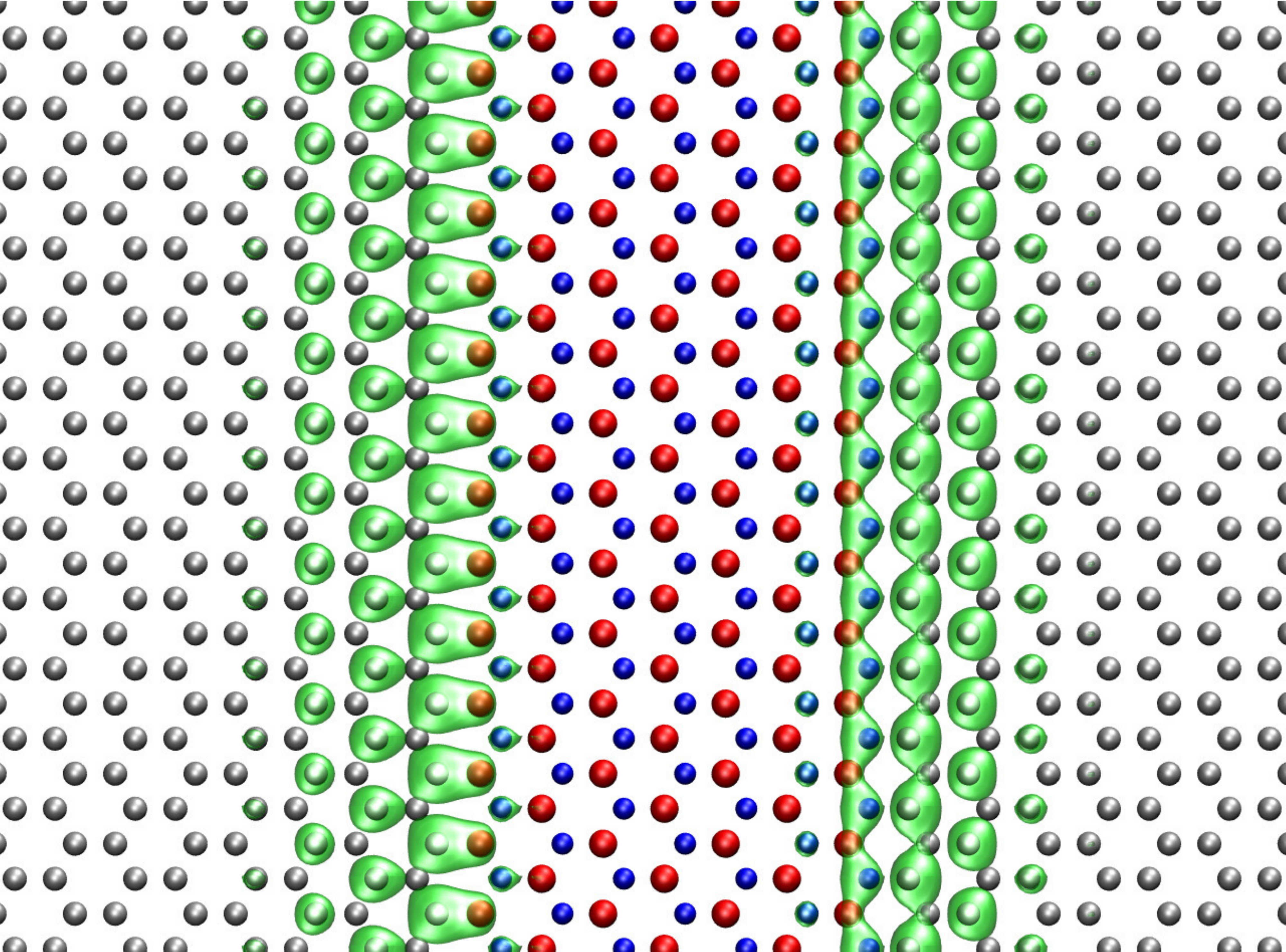}
\caption{Schematic of a h-BN nanoribbon embedded in graphene.  Two degenerate states localized at the N and B terminated zigzag edges of the h-BN nanoribbon are depicted by isosurfaces of $\pm$0.1$e$/\AA$^{3/2}$.  B, N, and C atoms are depicted by red, blue and grey balls, respectively.}\label{FigS8}
\end{figure}

\begin{figure}
\centering
\includegraphics[width=0.9\columnwidth]{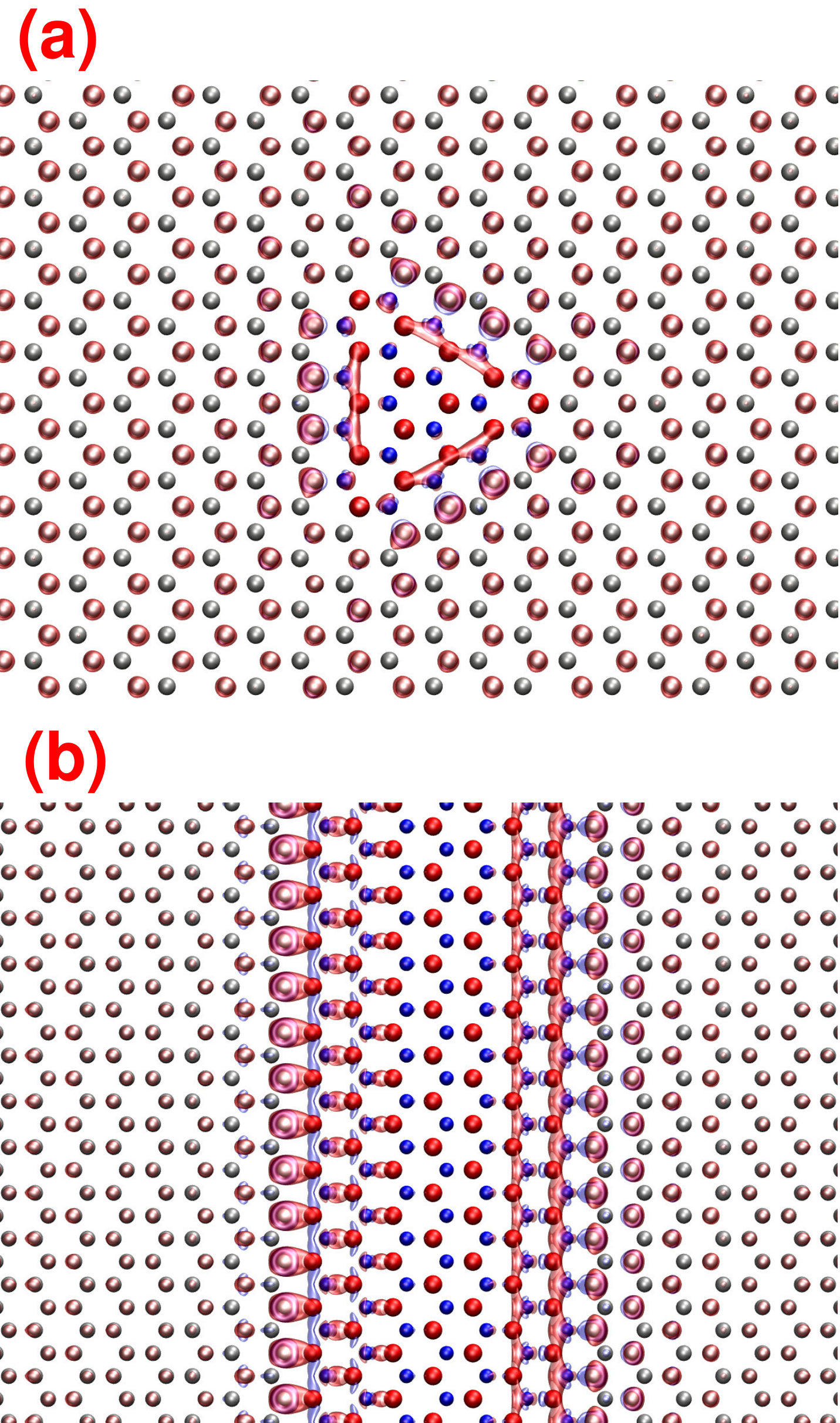}
\caption{Schematic of a (a) h-BN zigzag terminated nanoribbon and a (b) N terminated zigzag edged h-BN nanodomain embedded in graphene.  The change in charge density $\Delta\rho$ upon removal of one electron is shown by isosurfaces of $\pm$0.00625 $e$/\AA$^3$.  Red regions depict positive charge, while blue regions depict negative charge.  B, N, and C atoms are depicted by red, blue and grey balls, respectively.
}\label{FigS10}
\end{figure}

To better understand the underlying mechanisms behind the IMT in h-BNC systems, we have performed ab initio calculations at the density functional theory (DFT) level for several model h-BNC type systems.  These include h-BN nanoribbbons of various widths embedded in graphene (depicted schematically in Figure \ref{FigS8}), and h-BN nanodomains embedded in graphene (shown in Figure \ref{Fig4}
c and d).

Figures \ref{FigS10} and \ref{Fig4}c depict Kohn-Sham states localized on the edges of h-BN nanoribbons and nanodomains embedded in graphene, respectively.  For both systems, these states are near the Fermi level (at 0 and 0.2 eV respectively), suggesting that they may be the states through which VRH occurs.

\begin{figure}[b]
\centering
\includegraphics[width=\columnwidth]{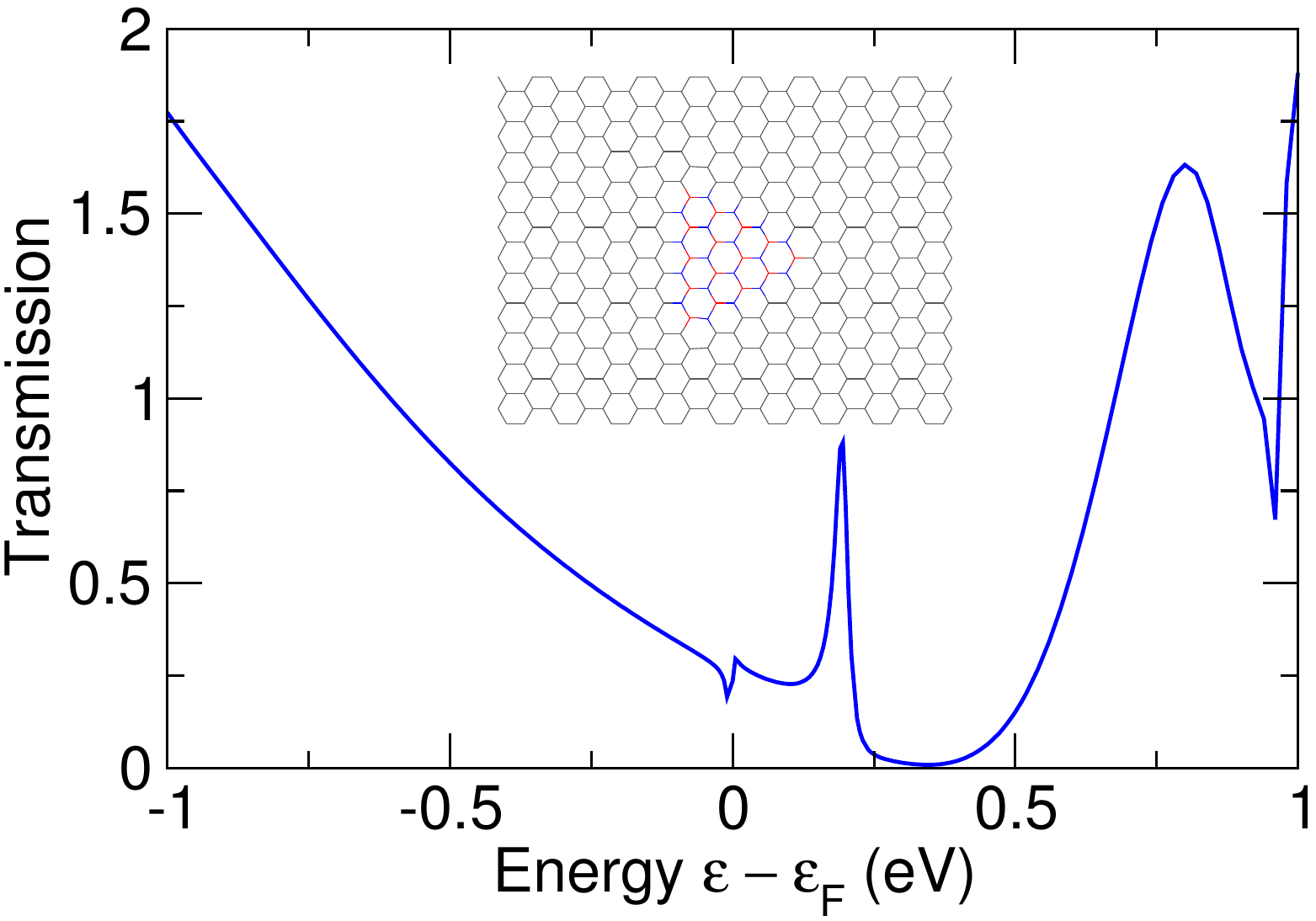}
\caption{Transmission through a N terminated zigzag edged h-BN nanodomain  embedded in graphene (shown as inset) as a function of energy $\varepsilon$ relative to the Fermi level $\varepsilon_F$ in eV.  Note the peak in the transmission at $\sim$ 0.2 eV corresponding to the state localized on the edge of the h-BN nanodomain, shown in Figure \ref{Fig4}
(c).
}\label{FigS9}
\end{figure}

Transport calculations have been performed using the non-equilibrium Green's function (NEGF) method with the electronic Hamiltonian obtained from the DFT calculation.  The calculated transmission through a N terminated zigzag edged h-BN nanodomain embedded in graphene is shown in Figure \ref{FigS9}, with a structural schematic as an inset.  The zero-bias transmission at the Fermi level is $\sim$ 0.2, giving a resistance of $\sim$ 60 k$\Omega$.  This agrees qualitatively with the low-temperature limit of resistance in the experimental system of ~ 200 k$\Omega$.  It should be noted that the peak in transmission at $\sim$ 0.2 eV corresponds to the position of the state localized on the N terminated zigzag edge of the h-BN nanodomain, shown in Figure \ref{Fig4}.  
Calculations of the transmission across h-BN nanoribbons embedded in graphene, shown schematically in Figure \ref{FigS8}, indicate these structures act as insulators, with resistances in the T$\Omega$ range, even for the smallest nanoribbon widths considered.

Investigations have also been performed into how doping via an external gate voltage influences the states in h-BNC systems.  This has been accomplished by performing charged calculations for both zigzag terminated h-BN nanoribbons and N-terminated zigzag edged h-BN nanodomains embedded in graphene.  In this case, a single valence electron has been removed from the supercell, and replaced by a uniform background electric potential to ensure no dipole-dipole interaction occurs between neighboring supercells in the periodic calculation.  Such calculations effectively mimic the effect of an external gate voltage (or doping) on the system.

Figure \ref{FigS10} depicts structural schematics and the change in charge density $\Delta\rho$ upon removal of one electron for h-BN zigzag terminated nanoribbons and N terminated zigzag edged h-BN nanodomains embedded in graphene.  While charge is removed from the extended graphene-like states in both systems, it is also transferred from the localized edge states on the h-BN nanoribbons and nanodomains to more tightly bound states also located on these edges.  More importantly, the energy of the localized edge states is nearly unchanged relative to the Fermi level.  This suggests that for the level of doping considered in the experiments discussed, these localized edge states are kept near the Fermi level, and available for VRH to occur.


\end{document}